\documentclass[12pt,epsf,amstex]{article}
\usepackage [dvips]{graphicx}
\usepackage{amsmath}
\usepackage{amssymb}
\usepackage{epsfig}
\usepackage{verbatim}

\addtocounter{secnumdepth}{1}
\begin{document}
\newcommand{\br}{\overline{r}}
\newcommand{\bbeta}{\bar{\beta}}
\newcommand{\bgamma}{\bar{\gamma}}
\newcommand{\tbeta}{\tilde{\beta}}
\newcommand{\tgamma}{\tilde{\gamma}}
\newcommand{\betap}{\beta^\prime}
\newcommand{\gammap}{\gamma^\prime}
\newcommand{\boldbeta}{\boldsymbol{\beta}}
\newcommand{\boldgamma}{\boldsymbol{\gamma}}
\newcommand{\calA}{{\cal A}}
\newcommand{\calO}{{\cal O}}

\newcommand{\bR}{{\bf{R}}}
\newcommand{\bS}{{\bf{S}}}
\newcommand{\half}{\frac{1}{2}}
\newcommand{\thalf}{{\textstyle{\frac{1}{2}}}}
\newcommand{\summ}{\sum_{m=1}^n}
\newcommand{\sumqno}{\sum_{q\neq 0}}
\newcommand{\tsum}{\Sigma}
\newcommand{\bsA}{\mathbf{A}}
\newcommand{\bsV}{\mathbf{V}}
\newcommand{\bsE}{\mathbf{E}}
\newcommand{\bsZ}{\hat{\mathbf{Z}}}
\newcommand{\bse}{\mbox{\bf{1}}}
\newcommand{\bspsi}{\hat{\boldsymbol{\psi}}}
\newcommand{\cdottt}{\!\cdot\!}
\newcommand{\deltaR}{\delta\mspace{-1.5mu}R}

\newcommand{\rmG}{\rm{G}}
\newcommand{\pnG}{p_n^{\rmG}}
\newcommand{\PnG}{P_n^{\rmG}}
\newcommand{\pnGz}{p_n^{\rmG,0}}
\newcommand{\PnGz}{P_n^{\rmG,0}}
\newcommand{\chiG}{\chi^{\rmG}}
\newcommand{\Rav }{\bar{R}}
\newcommand{\Ravc}{\bar{R}_{\rm c}}
\newcommand{\sigmav }{\bar{\sigma}}
\newcommand{\sigmavc}{\bar{\sigma}_{\rm c}}
\newcommand{\betaprime}{\beta^{\,\prime}}
\newcommand{\gammaprime}{\gamma^{\,\prime}}
\newcommand{\astar}{a^*}
\newcommand{\ap}{a^+}
\newcommand{\am}{a^-}
\newcommand{\tmu}{t}
\newcommand{\barnu}{\bar{\nu}}
\newcommand{\W}{U}

\newcommand{\bGamma}{\boldmath$\Gamma$\unboldmath}
\newcommand{\dd}{\mbox{d}}
\newcommand{\ee}{\mbox{e}}
\newcommand{\ii}{\mbox{i}}
\newcommand{\iif}{\mbox{\scriptsize i}}

\newcommand{\p}{\partial}

\newcommand{\la}{\langle}
\newcommand{\ra}{\rangle}
\newcommand{\rao}{\rangle\raisebox{-.5ex}{$\!{}_0$}}  
\newcommand{\rae}{\rangle\raisebox{-.5ex}{$\!{}_1$}}
\newcommand{\raG}{\rangle_{_{\!G}}}
\newcommand{\rainr}{\rangle_r^{\rm in}}
\newcommand{\beq}{\begin{equation}}
\newcommand{\eeq}{\end{equation}}
\newcommand{\bea}{\begin{eqnarray}}
\newcommand{\eea}{\end{eqnarray}}
\def\lsim{\:\raisebox{-0.5ex}{$\stackrel{\textstyle<}{\sim}$}\:}
\def\gsim{\:\raisebox{-0.5ex}{$\stackrel{\textstyle>}{\sim}$}\:}

\numberwithin{equation}{section}

\thispagestyle{empty}
\title{{\Large {\bf Many-sided Poisson-Voronoi cells\\[3mm] 
with only Gabriel neighbors\\
\phantom{xxx} }}}

\author{H.J. Hilhorst\\[3mm]
Laboratoire Ir\`ene Joliot-Curie, B\^atiment 210\\
Universit\'e Paris-Saclay, 91405 Orsay Cedex, France\\}
 
\maketitle
\begin{small}
\begin{abstract}
\noindent 

Let $p_n^G$ be the probability for a planar Poisson-Voronoi cell to be $n$-sided {\it and\,} have only Gabriel neighbors.
Using an exact coordinate transformation followed by scaling arguments and a mean-field type calculation, we obtain the asymptotic expansion of $\log\pnG$ in the limit of large $n$.
We determine several statistical properties of a many-sided cell obeying this `Gabriel condition.' In particular, the cell perimeter,
when parametrized as a function $\tau(\theta)$ of the polar angle $\theta$, behaves as a Brownian bridge on the interval $0\le\theta\le 2\pi$. 
We point out similarities and differences with related problems in random geometry.
\\

\end{abstract}
\end{small}
\vspace{5mm}

\newpage


\section{Introduction} 
\label{sec:introduction}
\vspace{5mm}

In the literature many questions have been asked about the statistics of Poisson-Voronoi tessellations. Scientists in a wide diversity of fields have investigated these random structures because of their numerous applications \cite{Okabeetal00}.
One question that has received strong interest from physicists \cite{Christetal82,ID89} is the following.
What is, in the two-dimensional plane, the probability $p_n$ that the cell of a randomly chosen seed have exactly $n$ sides?  The asymptotic large-$n$ behavior of $p_n$
was analyzed numerically and analytically more than forty years ago by Drouffe and Itzykson \cite{DI84}. Much later the present author \cite{Hilhorst05a,Hilhorst05b} returned to this question and obtained  the analytic result
\beq
p_n = 2C\,\frac{(8\pi^2)^{n-1}}{(2n)!} [ 1+o(1) ], \qquad n\to\infty,
\label{resultpn}
\eeq
where
$ C = \prod_{q=1}^\infty\,( 1-q^{-2}+4q^{-4} )^{-1} =  0.344\,347...$.
A Monte Carlo algorithm based on the derivation of (\ref{resultpn})
was developed \cite{Hilhorst07} for finding $p_n$ with high numerical precision for arbitrary $n$.
It was also shown that, in the limit of large $n$, 
the perimeter of an $n$-sided cell tends to a circle, and that the constant $C$ in (\ref{resultpn}) is due, in physical parlance, to the entropy of `elastic' deformations of the circle of wavelengths $2\pi n/q$.

The key initial step in deriving equation (\ref{resultpn})
is to cast the problem in terms of angular variables. This procedure is well adapted, more generally, to studying problems in two-dimensional random geometry that are statistically rotational invariant. The method was successfully used \cite{HilhorstCalka08} to study the Crofton cell (that is, the cell containing the origin) of a class of random line tessellations. It was also instrumental \cite{HilhorstCalkaSchehr08}
in establishing a link  between the answer to Sylvester's question \cite{Pfiefer89} and the Random Acceleration Process \cite{Burkhardt93,GMOR05}.
It was furthermore extended \cite{Hilhorst16} to the study of the $n$-sided face between two adjacent three-dimensional Poisson-Voronoi cells and to an analysis
\cite{HilhorstLazar14} of computer simulations initiated by
Mason {\it et al.} \cite{Masonetal12} and by Lazar {\it et al.} \cite{Lazaretal13}.
Further work along these lines \cite{Hilhorst09a,Hilhorst09b} has dealt with heuristic approximations
to the exact derivations as well as with extensions of those approximations to higher dimensions, where exact methods are lacking.
It has, moreover, led to a refutation \cite{Hilhorst06}
of Aboav's linear law \cite{Aboav70} as applied to planar Poisson-Voronoi cells,
and to a heuristic analysis \cite{Hilhorst14} of analogous laws in higher dimensions.\\

\begin{figure}
\begin{center}
\scalebox{.35}
         {\includegraphics{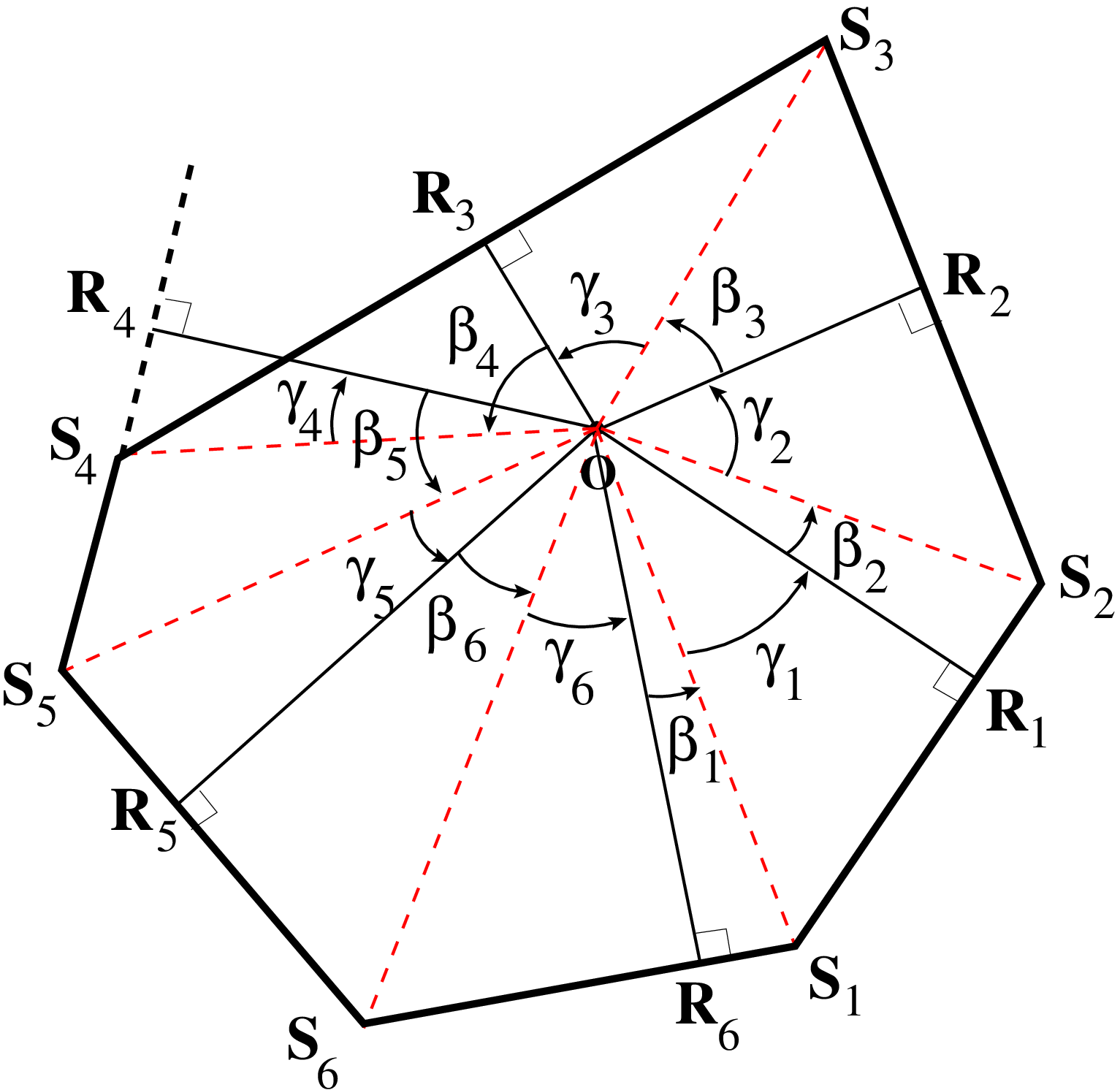}}
\end{center}
\caption{\small 
 Heavy black line: the perimeter of a six-sided Poisson-Voronoi cell having its seed in the origin ${\bf O}$.
 The seeds of the neighboring cells, not shown, are at positions
 $2\bR_m$, where $m=1,\ldots,n$.
 Thin solid black lines and thin dashed red lines connect the origin to the midpoints $\bR_m$ and the vertices $\bS_m$, respectively
The figure defines the angles $\beta_m$ as the increments of the polar angle when one advances along the perimeter from $\bR_{m-1}$ to $\bS_m$,  and the  $\gamma_m$ as the increments when one advances from $\bS_m$ to $\bR_m$.
The $m$th neighbor cell is a Gabriel neighbor if $\bR_m$ belongs to the perimeter segment $\overline{{\bS_m\bS_{m+1}}}$, and is not if $\bR_m$ belongs to its extension.
The latter situation occurs for $m=4$, with $\bR_4$ located on the extension of segment $\overline{{\bS_4\bS_5}}$, outside of the segment itself.
Hence this perimeter segment is not Gabriel; it is associated with a negative angle $\gamma_4$. 
For a general Poisson-Voronoi cell having its seed in ${\bf O}$ the $\beta_m$ and $\gamma_m$ may be of either sign; but it is easily seen that the cell has only Gabriel neighbors
if and only if the $\beta_m$ and $\gamma_m$ are positive for all $m$.}
\label{fig1}
\end{figure}

In this note we ask a new question that is in the same class
as the problems of references \cite{Hilhorst05b,HilhorstCalka08,HilhorstCalkaSchehr08,Hilhorst16}.
In a two-dimensional Poisson-Voronoi tessellation we consider a randomly selected cell (the `central cell') together with its neighbor cells.
Two neighboring cells are called `Gabriel neighbors' of one another if the line segment connecting their seeds intersects the {\it perimeter segment\,} that the two cells have in common, as opposed to intersecting the {\it extension of this perimeter segment}.
If the cells represent countries and the seeds their capitals, then two countries are Gabriel neighbors if the line segment connecting the two capitals does not pass through a third country.
Figure \ref{fig1} shows a six-sided Poisson-Voronoi cell that has five Gabriel neighbors and one non-Gabriel neighbor.
When a cell has only Gabriel neighbors we will say that it `satisfies
the Gabriel condition.'

Our question is: what is the probability, to be called $\pnG$,
that the central cell be $n$-sided {\it and\,} satisfy the Gabriel condition?
The main results of this work are the following.
We derive the asymptotic identity
\beq
\pnG = \frac{(16\pi^2)^{n-1}(n-1)!}{(15\pi n^3)^{1/2}(3n-3)!}
\,[ 1+\calO(n^{-1}) ],\qquad n\to\infty,
\label{resultpnG}
\eeq
which with the aid of Stirling's formula may be written alternatively as
\beq
\log\pnG = -2n\log n + n\log(2^4 3^{-3}\pi^2\ee^2)
+ \thalf\log(2^{-8}3^{4}5^{-1}\pi^{-5} n) + \calO(n^{-1}).
\label{resultlogpnG} 
\eeq
The ratio $\phi_n \equiv p_n^{\rmG}/p_n$ is the fraction of $n$-sided Voronoi cells that have only Gabriel neighbors.
From (\ref{resultpn}) and (\ref{resultpnG}) we find
\beq
\phi_n = \frac{3^2 n}{2\cdot 5^{1/2}\,C}\,
\left(\frac{2}{3}\right)^{\!3n} [ 1+o(1) ], \qquad n\to\infty.
\label{resultphin}
\eeq
We show that the $n$-sided cell satisfying the Gabriel condition
has a perimeter that for large $n$ tends towards a circle of radius $(n/4\pi)^{1/2}+\calO(1)$,
where the $\calO(1)$ term is a random contribution, and
that the cell perimeter runs mostly through a narrow  annulus of width of $\calO(n^{-1})$.
This means that the Gabriel condition renders the central cell more rigidly circular than it would have been without this condition. Concomitantly, contributions to (\ref{resultpnG}) from elastic deformations appear only in the $\calO(n^{-1})$ terms of the asymptotic expansion.

In section \ref{sec:basicintegral} we start from the basic expression for $\pnG$, which is a $2n$-dimensional integral, and recall concisely how to transform it to an integration over only angular variables.
In section \ref{sec:largen} 
we perform a large-$n$ expansion of the transformed expression.
This leads to a mean-field type problem which we identify and solve in section \ref{sec:meanfield} and which yields (\ref{resultpnG}). 
In section \ref{sec:distrcell} we analyze the probability distribution of the angular variables and the shape of the cell perimeter.
In the limit $n\to\infty$ the latter appears to satisfy a simple Langevin equation.
In section \ref{sec:conclusion} we present some further comments and conclude. 

\section{$\pnG$ as a $2n$-dimensional integral}
\label{sec:basicintegral}

We consider an infinite plane in which seeds are distributed uniformly and independently with unit density.%
\footnote{In the mathematical literature the distribution of the seeds is called a {\it Poisson process} and their density is referred to as an
  {\it intensity}.}
We select a seed at random, choose it as the origin of the coordinate system, and ask for the probability $\pnG$ that its Voronoi cell (the `central cell') have exactly $n$ neighboring cells which, moreover, are all Gabriel neighbors.
The perimeter of the central cell is completely specified by the positions
$2\bR_m$, with $m=1,2,\ldots,n$, of the seeds of the $n$ neighboring cells.
The perpendicular bisector of the vector $2\bR_m$ passes through the
midpoint $\bR_m$ (see figure \ref{fig1}) and contains the $m$th segment of the perimeter of the central cell. This midpoint may lie on that segment itself,
or it may lie on that segment's extension.
In the first case the central cell and its $m$th neighbor are Gabriel neighbors
(we will say for short that the $m$th perimeter segment `is Gabriel'), whereas in the second case they are not.


\subsection{Expression for $\pnG$ in Cartesian coordinates}
\label{sec:cartesian}

The definition of $\pnG$ leads directly to the $2n$-dimensional `phase space' integral
\begin{equation}
\pnG = \frac{4^n}{n!}
\int \dd\bR_1 \ldots \dd\bR_n\,\,
\chiG(\bR_1,\ldots,\bR_n)\,\, 
\ee^{-4{\cal A}(\bR_1,\ldots,\bR_n)},
\label{pnG1}
\end{equation}
in which

(i) $\chiG$ is the indicator function for the domain in phase space 
for which the $n$-tuplet $\{\bR_1,\ldots,\bR_n\}$ defines a 
convex $n$-sided polygon all whose perimeter segments are Gabriel.
convention 4*lambda=1; at the end of the calculation lambd mau be restored. 

(ii) $4{\cal A}$ is the area of a region surrounding the central seed,
often called its {\it flower,} that should be free of seeds other than those at $2\bR_1,\ldots,2\bR_n$ lest the condition of the cell having $n$ neighbors
be violated.
The exponential $\ee^{-{4\cal A}}$ is the probability that this condition is respected.

Explicit expressions for ${\cal A}$ and $\chi^G$ 
may be found with the aid of basic but tedious trigonometry;
several authors have recast them in various different forms
\cite{DI84,Calka02,CalkaSchreiber05,Hilhorst05b}.
The integral (\ref{pnG1}) is reminiscent of the partition function
of $n$ interacting two-dimensional particles.


\subsection{Transforming $\pnG$  to angular coordinates}
\label{sec:angular}

We will apply to
expression (\ref{pnG1}) a highly nontrivial transformation
that has been described in detail elsewhere \cite{Hilhorst05b} and that we will therefore state here only concisely.
We introduce the set of angles $\{\beta_1,\gamma_1,\ldots,\beta_n,\gamma_n\}$ defined in figure \ref{fig1} and which
satisfy the sum rule 
\beq
\sum_{m=1}^{n}(\beta_m + \gamma_m) = 2\pi.
\label{sumrule1}
\eeq
These angles define the central cell up to an overall rotation and a homothety with respect to the origin. Rotating the cell contributes a factor $2\pi$ to $\pnG$. The cell's radial scale may be fixed by a single 
radial variable, for which we take the mean $\Rav$ of the `radii' $R_m\equiv|\bR_m|$, that is, 
\beq
\Rav = n^{-1}\sum_{m=1}^nR_m\,.
\label{dRav}
\eeq
We will refer to $\Rav$ as the cell's mean radius.
Furthermore we set
\beq
\calA \equiv \pi\Rav^2 \big( 1+n^{-2}\W(\boldbeta,\boldgamma) \big),
\label{defU}
\eeq
which defines $\W$.
The integration over $\Rav$ decouples from the angular integrations and is easily carried out.%
\footnote{We return to this integral in section \ref{sec:circularity}.}
We find, after a considerable amount of trigonometry, that the initial integral (\ref{pnG1}) may be alternatively expressed as
\bea
\pnG = \frac{(n-1)!}{2n\pi^n} \int\!\dd\boldbeta\dd\boldgamma\,\,
\delta\big( G \big)\,
\delta\Big( \sum_{m=1}^n(\beta_m + \gamma_m)-2\pi \Big)\,
\left[ \prod_{m=1}^n(\beta_m+\gamma_m) \right]\,\ee^{-\mathbb V}
\nonumber\\
&&
\label{pnG2}
\eea
in which
\beq
\ee^{-\mathbb V} = 
\left[ \prod_{m=1}^n \frac{\rho_m^2 \sin(\beta_m+\gamma_m)}
  {(\beta_m+\gamma_m)\cos\beta_m\cos\gamma_m} \right]\,
\big( 1+n^{-2}\W \big)^{-n}
\label{deVV}
\eeq
and where we have abbreviated
\bea
\int\dd\boldbeta\,\dd\boldgamma 
= \int_0^{\pi/2} \prod_{m=1}^n \dd\beta_m\dd\gamma_m\,.
\label{defintbetagamma}
\eea
When (\ref{deVV}) is substituted in (\ref{pnG2})
a factor $\prod_{m=1}^n(\beta_m+\gamma_m)$ cancels out; the splitup as stated between equations (\ref{deVV}) and (\ref{pnG2}) will nevertheless turn out to have an essential advantage in section \ref{sec:largen}.

The symbols $G, \rho_m$, and $\W$ in (\ref{pnG2})-(\ref{deVV}) have the following meaning.
The `reduced radii' $\rho_m\equiv R_m/\Rav$
may be expressed in terms of the $\beta_m$ and $\gamma_m$ by
\beq
\rho_m = \left[ \prod_{\ell=1}^m\frac{\cos\gamma_\ell}{\cos\beta_\ell} \right]\,
\rho_n\,, \qquad m=1,\ldots,n-1,
\label{rhocosbcosg}
\eeq
and the sum rule $n^{-1}\sum_{m=1}^n\rho_m=1$.
The geometry dictates that $\rho_m$ be $n$-periodic in the index $m$, that is, equation (\ref{rhocosbcosg}) should hold also for $m=n$.
This leads to the appearance in (\ref{pnG2}) of the factor $\delta(G)$, where
\beq
G(\boldbeta,\boldgamma) = \frac{1}{2\pi} 
\log \prod_{m=1}^n\frac{\cos\gamma_m}{\cos\beta_m}\,.
\label{defG}
\eeq
Finding $\W(\beta,\gamma)$ in terms of the angular variables requires a lengthy calculation that leads to
\bea
\W(\boldbeta,\boldgamma) &=& 
\frac{n^2}{4\pi}\summ \rho_m^2 \big[ \tan\gamma_m-\gamma_m 
+\tan\beta_{m+1}-\beta_{m+1} \nonumber\\
&& \phantom{xxxxxxxxx} + \gamma_{m}\tan^2\gamma_{m}
   + \beta_{m+1}\tan^2\beta_{m+1} \big] \nonumber\\
&& +\, \frac{n^2}{2\pi}
   \summ (\rho_m^2-1)(\gamma_m + \beta_{m+1})
\label{xVbetagamma}
\eea
with the convention that $\beta_{n+1}=\beta_1$.
All quantities occurring in the integrand in (\ref{pnG2}) may now be expressed as functions of the angles $\beta_m$ and $\gamma_m$
by means of the relations of this section.
We stress that the transformation
that takes us from (\ref{pnG1}) to (\ref{pnG2}) is fully exact
and is valid for all $n=3,4,\ldots$. 

We still observe that in passing from (\ref{pnG1}) to
(\ref{pnG2}) we have lost the function $\chi^G$:
in fact, it has left its mark on the {\it range\,} of the angular variables in (\ref{defintbetagamma}), which are all integrated from $0$ to $\pi/2$.
These limits of integration differ from those in the analogous problem without the Gabriel condition \cite{Hilhorst05b}, and
this seemingly minor difference will require a distinctly different analysis in what follows.


\section{Large-$n$ expansion}
\label{sec:largen}

We here set out to extract from the $2n$-fold integral (\ref{pnG2}) its asymptotic behavior for large $n$, using formal expansions in powers of $n^{-1}$ and a variety of scaling arguments that are common in theoretical physics and generally lead to exact results.


\subsection{Scaling of the angles $\beta_m$ and $\gamma_m$}
\label{sec:scbg}

In view of the positivity of the $\beta_m$ and $\gamma_m$,
the symmetry between
them, and the sum rule (\ref{sumrule1}), these angles must scale with $n$ as
\beq
\beta_m\,, \gamma_m \sim n^{-1}.
\label{scbg}
\eeq
Here and below the symbol $\sim$ indicates how quantities scale with $n$.
The power of $n$ in (\ref{scbg}) is different from the one that occurs in the problem without the Gabriel condition \cite{Hilhorst05b}, where $\beta_m\,, \gamma_m \sim n^{-1/2}$. This difference will impact all that follows.


\subsection{The factor $\ee^{-{\mathbb V}}$ for large $n$}
\label{sec:expVV}

From here on we proceed on the basis of a hypothesis
that will be confirmed self-consistently by the calculations based on it.
It holds that as $n\to\infty$ the pairs of angles $(\beta_m,\gamma_m)$, for $m=1,2,\ldots,n$ become $n$ mutually independent two-dimensional stochastic variables.
This hypothesis will first and foremost lead to the key relation
\beq
\ee^{-\mathbb V} = 1 + \calO(n^{-1}),
\label{scexpmV} 
\eeq
which is going to considerably simplify all further work.

In order to arrive at (\ref{scexpmV}) we will consider
successively the factors on the RHS of equation (\ref{deVV})
that make up $\ee^{-\mathbb V}$.
We let $\tau_m \equiv \rho_m - 1$ so that the sum rule on the $\rho_m$ stated below (\ref{rhocosbcosg}) translates into
\beq
\sum_{m=1}^n\tau_m=0.
\label{sumrule2}
\eeq
Then, first of all, expanding relation (\ref{rhocosbcosg}) for small $\beta_\ell$ and $\gamma_\ell$ yields
\bea
\tau_m -\tau_0 &=& \frac{1}{2}\sum_{\ell=1}^m\Big( \beta_\ell^2 - \gamma_\ell^2 +
\calO(n^{-4})\Big)
\label{expdifftau}
\eea
with the convention $\tau_0=\tau_n$.
If we now use that to leading order in $n$ we may
consider $\tau_m-\tau_0$ as a sum of $m$ independent identically distributed zero average random variables $\thalf(\beta_\ell^2-\gamma_\ell^2)$, each one distributed on scale $n^{-2}$, and take $m\sim n$, then equation (\ref{expdifftau}) leads to
$\tau_m-\tau_0 \sim n^{-3/2}$.
When combined with the sum rule (\ref{sumrule2}) this leads to
\beq
\tau_m\sim n^{-3/2}.
\label{sctau}
\eeq
At this point it suffices that we know the scaling (\ref{sctau}); in section \ref{sec:circularity} more quantitative conclusions will be drawn from equation (\ref{expdifftau}).

Secondly, taking into account sum rule (\ref{sumrule2}) we see that
\bea
\log\prod_{m=1}^n\rho_m^2 &=& -\sum_{m=1}^n \left( \tau_m^2 + \calO(\tau_m^3)
\right)
\nonumber\\[2mm]
&\sim& n\cdot (n^{-3/2})^2 \nonumber\\[2mm]
&\sim& n^{-2},
\label{scprodrho}
\eea
where, to pass to the second line, we used the scaling (\ref{sctau}).

Third, straightforward expansion for small $\beta_m$ and $\gamma_m$ yields
\bea
\log\prod_{m=1}^n\frac{\sin(\beta_m+\gamma_m)}
                {(\beta_m+\gamma_m)\cos\beta_m\cos\gamma_m)} &\sim& n^{-1}.
\label{scT}
\eea

Fourth, the quantity $\W$ of equation (\ref{xVbetagamma}) is composed of two sums that we will call $\W_1$ and $\W_2$ and consider separately.
When expanding $\W_1$ for small $\beta_m$, $\gamma_m$, and $\tau_m$ we find
\bea
\W_1 &=& \frac{n^2}{6\pi}\sum_{m=1}^n\big( \beta_m^3 + \gamma_m^3 +
\calO(n^{-5}) \big)
\nonumber\\[2mm]
&\sim& n \cdot n^2 \cdot (n^{-3/2})^2 \nonumber\\[2mm]
&\sim& 1.
\label{scV1}
\eea
In order to discuss the term $\W_2$ we write $\gamma_m=\pi n^{-1}+\delta\gamma_m$ and $\beta_m=\pi n^{-1}+\delta\beta_m$,
so that now the $\delta\beta_m$ for $m=1,2,\ldots,m$ are independent zero average random variables of order $\calO(n^{-1})$; and so are the $\delta\gamma_m$. We then have
\bea
\W_2 &=& \frac{n^2}{2\pi}\sum_{m=1}^n \big[ 2\tau_m+\calO(\tau_m^2) \big]
[\gamma_m+\beta_{m+1}]
\nonumber\\[2mm]
&=& \frac{n^2}{\pi} \sum_{m=1}^n \big[ \tau_m +\calO(\tau_m^2) \big]
    [2\pi n^{-1} + \delta\gamma_m + \delta\beta_{m+1}]\nonumber\\[2mm]
           &\sim& n^2\cdot n^{1/2}\cdot n^{-3/2}\cdot n^{-1}\nonumber\\[2mm]
           &\sim& 1.
\label{scV2}
\eea
To pass from the second to the third line we have reasoned as follows.
The cross term $2\pi n^{-1}\tau_m$, when summed on $m$, vanishes due to the sum rule (\ref{sumrule2}). The $\calO(\tau_m^2)$ term contributes only to higher order.
There remains only the cross term $\tau_m[\delta\gamma_m+\delta\beta_{m+1}]$,
where we have that $\tau_m=\calO(n^{-3/2})$,
and in view of what was said above about the $\delta\beta_m$ and $\delta\gamma_m$,
the sum on $m$ increases the order in $n$ by a factor $n^{1/2}$.

Since $\W=\W_1+\W_2$ we find that $\W$ is $\calO(1)$, which we anticipated
in our definition (\ref{defU}).
Upon employing relations
(\ref{scprodrho})-(\ref{scV2}) in (\ref{deVV}) we obtain
$\ee^{-\mathbb V} = 1 + \calO(n^{-1})$
as announced in (\ref{scexpmV}).


\section{A mean-field problem}
\label{sec:meanfield}

When in expression (\ref{pnG2}) we neglect the $\calO(n^{-1})$ terms that we identified so far,
the integrand factorizes into $n$ factors each depending on a single pair of variables $(\beta_m,\gamma_m)$, this apart from two $\delta$-function constraints,
one linear and the other one nonlinear in the $\beta_m$ and $\gamma_m$.
The constraints are global and 
the correlations that they create between different pairs of variables will vanish for $n\to\infty$.
Hence our initial hypothesis of these variables being independent in the limit $n\to\infty$ receives at this point its {\it a posteriori\,} confirmation.
The calculation of $\pnG$ has thus been reduced to a problem of mean-\-field
type that we will solve in the remainder of this section.
We will deal with the two constraints by introducing appropriate integral representations. 


\subsection{Transformation to variables $a_m^\pm$}
\label{sec:evaluation}

\noindent We define the linear combinations
\bea
\ap_m &=& \phantom{\thalf}(\beta_m + \gamma_m)n,  \nonumber\\[2mm]
\am_m &=&          \thalf (\beta_m - \gamma_m)n,  \qquad m=1,2,\ldots,n,
\label{dapm}
\eea
which are such that in the limit $n\to\infty$ the integrations become
\beq
\int_0^{\pi/2}\dd\beta_m\, \int_0^{\pi/2}\dd\gamma_m =
n^{-2} \int_0^\infty \dd\ap_m\, \int_{-\ap_m/2}^{\ap_m/2}\,\dd\am_m + \ldots,
\label{intapm}
\eeq
Here the dots stand for correction terms that must be expected to vanish exponentially with $n$ for the integrands that will be of interest here.
      In terms of the $a_m^{\pm}$ the quantity $G$ of equation (\ref{defG})
takes the form
\bea
G &=& \frac{1}{2\pi}
\sum_{m=1}^n \left[ \log\cos\gamma_m - \log\cos\beta_m \right]
\nonumber\\[2mm]
&=& \frac{1}{2\pi} \sum_{m=1}^n \left[ \thalf\beta_m^2 - \thalf\gamma_m^2
  + \calO(n^{-4}) \right] \nonumber\\[2mm]
&=& \frac{1}{2\pi n^2} \sum_{m=1}^n \left[ \ap_m\am_m + \calO(n^{-2}) \right].
\label{xG}
\eea
%
When using (\ref{scexpmV}) as well as (\ref{dapm})-(\ref{xG}) in (\ref{pnG2}) we obtain
\bea
\pnG &=& \frac{(n-1)!}{2n\pi^n} \cdot 2\pi n^2 \cdot n^{-3n+1}
\int\!\dd\ap_1\dd\am_1\,\,\dd\ap_2\dd\am_2\,
\ldots\,\dd\ap_n\dd\am_n\, \nonumber\\[2mm]
&& \times\, \delta\left( \sum_{m=1}^n \ap_m\am_m \right)
\delta\left( \sum_{m=1}^n\ap_m - 2\pi n \right)\,
\left[ \prod_{m=1}^n\ap_m \right]
\left[ 1+\calO(n^{-1}) \right],\nonumber\\
&&
\label{pnG5}
\eea
where the limits of integration on the $a_m^\pm$ are as in (\ref{intapm}).


\subsection{Two integral representations}
\label{sec:aspteval}

In order to evaluate (\ref{pnG5}) we introduce the integral representations
\beq
\delta\left( \sum_{m=1}^n \ap_m\am_m  \right) =
\frac{n^{-1/2}}{2\pi}
\int_{-\infty}^{\infty}\!\dd\tmu\, \ee^{\iif\tmu n^{-1/2} \sum_{m=1}^n \ap_m\am_m}
\label{irdeltaspir}
\eeq
and
\beq
\delta\left( \sum_{m=1}^n\ap_m - 2\pi n \right) = \frac{1}{2\pi\ii}
\int_{c-\iif\infty}^{c+\iif\infty}\!\dd s\, \ee^{2\pi sn-s\sum_{m=1}^n \ap_m}
\label{irdeltageom}
\eeq
where we take $c>0$.
The scaling factor $n^{-1/2}$ in (\ref{irdeltaspir}), which at this stage could have been any other power of $n$, will be essential below.
When we substitute (\ref{irdeltaspir}) and (\ref{irdeltageom}) in (\ref{pnG5}), the integrals on the $a_m^\pm$ decouple into $n$ identical pairs and we get 
\beq
\pnG = \frac{(n-1)!}{2n\pi^n}\cdot\frac{n^{5/2}}{n^{3n}}\cdot
\frac{1}{2\pi\ii}\int_{c-\iif\infty}^{c+\iif\infty}\!\dd s\,\ee^{2\pi sn}
\int_{-\infty}^{\infty}\!\dd\tmu\,I^{\,n}(\tmu,s)
\,\big[ 1 + \calO(n^{-1}) \big]
\label{pnG6}
\eeq
in which 
\bea
I(\tmu,s) &=& \int_0^\infty\!\dd\ap\,\ap\,\ee^{-s\ap}
\int_{-\ap/2}^{\ap/2}\!\dd\am\,\ee^{\iif\tmu n^{-1/2}\ap\am}\nonumber\\[2mm]
&=& \int_0^\infty\!\dd\ap\,\ee^{-s\ap}\left[ (\ap)^2 - \frac{t^2}{4!\,n}(\ap)^6
  + \calO(n^{-2}) \right] \nonumber\\[2mm]
&=& \frac{2}{s^3}\left[ 1 - \frac{15\tmu^2}{s^4n} + \calO(n^{-2})  \right].
\label{defImus}
\eea
The positivity of $c$ ensures the convergence of the $\ap$ integral in (\ref{defImus}).
Furthermore the scaling of $t$ with $n^{-1/2}$ adopted in equation (\ref{irdeltaspir}) now leads to a nontrivial $t$ dependence 
of $I^{\,n}(\tmu,s)$ when $n$ gets large. Explicitly,
\beq
I^{\,n}(\tmu,s) = \frac{2^n}{s^{3n}}\,\exp\left( -\frac{15\tmu^2}{s^4} \right)
\big[ 1 + \calO(n^{-1}) \big],
\label{limImus}
\eeq
whence
\beq
\int_{-\infty}^{\infty}\!\dd\tmu\,I^{\,n}(\tmu,s) =
\frac{2^n}{s^{3n-2}} \left( \frac{\pi}{15} \right)^{1/2}
\left[ 1+\calO(n^{-1})\right]. 
\label{intImus}
\eeq
We substitute (\ref{intImus}) in (\ref{pnG6}) and perform the integration on $s$ by letting its path run counterclockwise around the origin and employing the identity 
\beq
\frac{1}{2\pi\ii}\oint \dd s \frac{ \ee^{-2\pi sn} }{ s^{3n-2} } =
\frac{ (2\pi n)^{3n-3} }{ (3n-3)! }\,, \qquad n=1,2,\ldots.
\label{ints}
\eeq
This leads to
\beq
\pnG = \frac{(16\pi^2)^{n-1}}{(15\pi n^3)^{1/2}} \,          \frac{(n-1)!}{(3n-3)!}\,\left[ 1+\calO(n^{-1}) \right],
\label{pnG7}
\eeq
which is result (\ref{resultpnG}) announced in the introduction.\\


\section{Probability distributions, cell shape, and Langevin equation} 
\label{sec:distrcell}

In this section we are interested in the probability distributions of individual angles $\beta_m$ and $\gamma_m$.
For an arbitrary function $X(\boldbeta,\boldgamma)$
we let $\PnG[X]$ denote the same integral as $\pnG$ but with an insertion $X$ in the integrand; hence $\PnG[1]=\pnG$.
The average $\la X\ra$ is given by
\beq
\la X \ra = \frac{\PnG[X]}{\pnG},
\label{davX}
\eeq
which we will use below.


\subsection{The probability distribution $w_2(\beta_m,\gamma_m)$}
\label{sec:distributions}

We now ask about the probability distribution $w_2(\beta_m,\gamma_m)$ of the pair $(\beta_m,\gamma_m)$ in an $n$-sided Poisson-Voronoi cell with only Gabriel neighbors. We have 
\bea
w_2(\beta,\gamma) &=& \big\la \delta(\beta_m-\beta)\,\delta(\gamma_m-\gamma) \big\ra \nonumber\\[2mm]
&=& n^2 \big\la \delta(\thalf\ap_m + \am_m-\beta n)\,
\delta(\thalf\ap_m - \am_m-\gamma n) \big\ra.
\label{dw2}
\eea
We may implement the angle brackets in the second line of (\ref{dw2}) by directly using expression (\ref{pnG5}) and going through the same steps as in section \ref{sec:aspteval}. 
This leads to          
\beq
w_2(\beta,\gamma) = \frac{n^2}{2} \left( \frac{3}{2\pi} \right)^{\!\!3}
(\beta n+ \gamma n)\,\exp\left( - \frac{3(\beta n+\gamma n)}{2\pi} \right)
\big[ 1+\calO(n^{-1}) \big],
\label{xw2}
\eeq
valid in the limit of large $n$ with $\beta n$ and $\gamma n$ fixed, this latter condition being coherent with the scaling (\ref{scbg}).
Integration of (\ref{xw2}) over $\gamma$ yields the probability distribution $w_1(\beta_m)$ of the single variable $\beta_m$,
\beq
w_1(\beta) = \frac{3n}{4\pi}\left( 1+\frac{3\beta n}{2\pi} \right)
\exp\left( -\frac{3\beta n}{2\pi} \right)
\big[ 1+\calO(n^{-1}) \big],
\label{xw1}
\eeq
valid in the limit of large $n$ with $\beta n$ fixed.
By symmetry (\ref{xw1}) is also the distribution of $\gamma_m$.
It correctly reproduces $\la\beta_m\ra = \la\gamma_m\ra = \pi/n$.

In the limit $n\to\infty$ the effect of the two global constraints
on locally defined quantities will vanish. In particular, for $n\to\infty$ the pairs $(\beta_m,\gamma_m)$ with $m=1,2,\ldots,n$
become $n$ independent two-dimensional random variables.
This structure of the angular statistics is distinct from that found in earlier work \cite{Hilhorst05b} in the absence of the Gabriel condition.
In that case there exist
$2n$ asymptotically independent angles of two kinds, namely\,
$\xi_m \equiv \beta_m + \gamma_m$ and $\eta_m \equiv \gamma_m + \beta_{m+1}$,
with distributions that also involve linear terms and exponentials, but are different from (\ref{xw2}) and (\ref{xw1}).


\subsection{Cell shape}
\label{sec:circularity}

We will show that for large $n$ the cell perimeter of the Poisson-Voronoi cell satisfying the Gabriel condition tends towards a circle and quantify the deviations from circularity. We abbreviate
\beq
\epsilon_m \equiv \thalf(\beta_m^2 - \gamma_m^2), \qquad m=1,2,\ldots,n.
\label{ddeltam}
\eeq
The variance $\la\epsilon_m^2\ra$ under the distribution $w_2(\beta,\gamma)$  
is readily calculated with the result
\beq
\la\epsilon_m^2\ra = \frac{30}{n^4}\left(\frac{2\pi}{3}\right)^{\!\!4}
\big[ 1+\calO(n^{-1}) \big].
\label{xdeltam2}
\eeq

We will also need the correlation
$\la\epsilon_\ell\epsilon_m\ra$ where $\ell\neq m$.
Because of the mean-field character of this problem
this correlation is independent of $\ell$ and $m$.
To leading order in $n$ equation (\ref{sumrule2})
implies that $\sum_{m=1}^n \epsilon_m = 0$.
This leads to $\big\la \left( \sum_{m=1}^n \epsilon_m \right)^2 \big\ra = 0$,
whence
\beq
\la\epsilon_m\epsilon_{m+\ell}\ra = \la\epsilon_m^2\ra\, 
\left[ \frac{n}{n-1} \delta_{\ell,0}
-\frac{1}{n-1} \right], 
\label{relml}
\eeq
which we will use below.\\

The integration over $\Rav$ mentioned in section (\ref{sec:angular})
reads, explicitly,
$\int_0^\infty\dd\Rav\,$ $\Rav^{2n-1} 
\exp\left( -4\pi\Rav^2[ 1+n^{-2}U(\boldbeta,\boldgamma)] \right)$.
For large $n$ the integrand gets strongly peaked around
$\Rav=\Ravc$ with a standard deviation $\sigmav$.
Both $\Ravc$ and $\sigmav$ are to leading order in $n$ independent of $(\boldbeta,\boldgamma)$ and are given by
\beq
\Ravc   = (n/4\pi)^{1/2}[ 1+\calO(n^{-1}) ], \qquad
\sigmav = (16\pi)^{-1/2}[ 1+\calO(n^{-1}) ].
\label{intRav}
\eeq
They describe averages taken over the {\it ensemble\,} of Poisson-Voronoi cells.
The mean radius $\Rav$ of an individual cell will have a random value
$\Rav = \Ravc + \calO(\sigmav) = \Ravc + \calO(1)$.
We now ask the finer question of how close the perimeter of an
{\it individual\,} cell approaches a circle when $n$ gets large.
A good characterization of the degree of approach is provided by the set of
averaged square differences $d_\ell^{\,2}$, for $\ell=0,1,\ldots,n$,
between the radii $R_m$ and $R_{m+\ell}$,
\beq
d_\ell^{\,2} = \langle (R_{m+\ell} - R_m)^2 \rangle.
\label{ddell}
\eeq
Because of translational symmetry in the index $m$, the $d_\ell^{\,2}$
cannot depend on $m$. We use that
\beq
R_{m+\ell}-R_m = \Rav (\tau_{m+\ell}-\tau_m).  
\label{xRml}
\eeq
In (\ref{xRml}) we replace $\Rav$ by its leading term, which is the constant $(n/4\pi)^{1/2}$. We then square this expression, substitute it in (\ref{ddell}), and use (\ref{expdifftau}) while employing the notation (\ref{ddeltam}).
The result is
\bea
d_\ell^{\,2} &=& \frac{n}{4\pi}\,
\big\la (\epsilon_{m+1}+\epsilon_{m+2}+\ldots+\epsilon_{m+\ell})^2 \big\ra
\nonumber\\[2mm]
&=& \frac{n}{4\pi}\,\left[ \ell\la\epsilon_m^2\ra + \ell(\ell-1)
 \la\epsilon_m\epsilon_{m+\ell}\ra \right] \nonumber\\[2mm]
&=& \frac{\ell(n-\ell)}{4\pi}\la\epsilon_m^2\ra\, [ 1+\calO(n^{-1} ],
\qquad \ell=0,1,\ldots,n, 
\label{xdell}
\eea
where to arrive at the last line we have used (\ref{relml}).
The proportionality of the RHS of (\ref{xdell}) to $\ell(n-\ell)$ is the well-known parabolic behavior of the mean square displacement of a Brownian bridge on an interval $[0,n]$.
Since $\la \epsilon_m^2 \ra = \calO(n^{-4}) $, equation (\ref{xdell})
tells us that for a typical value $\ell=\calO(n)$ we will have that
$d_\ell = \calO(n^{-1}) $. This means that the perimeter of a typical $n$-sided Poisson-Voronoi cell satisfying the Gabriel condition
runs through an annulus of radius $(n/4\pi)^{1/2}[1+\calO(n^{-1/2})]$
and of width $\calO(n^{-1})$. Without the Gabriel condition \cite{Hilhorst05b} the radius of an $n$-sided cell is the same  but the width of the annulus is  $\calO(1)$.
Loosely speaking, in the large $n$ limit a Poisson-Voronoi cell having only Gabriel neighbors deviates much less from a circle than a cell without this condition. This section has proven and quantified this intuitively appealing result.


\subsection{Langevin equation}
\label{sec:langevin}

We define the polar angle $\theta = 2\pi m/n$ and set
$\tau(\theta) = \tau_m$. In the limit of large $n$ the recursion
$\tau_{m}-\tau_{m-1} = \epsilon_m$
becomes the Langevin equation
\beq
\frac{\dd\tau}{\dd\theta} = \epsilon(\theta)
\label{rectautheta}
\eeq
in which $\epsilon(\theta)$ is zero average Gaussian noise whose covariance should be of the form
\beq
\langle\epsilon(\theta)\epsilon(\theta^\prime)\rangle =
n^{-3}D \left[ \,2\pi\delta(\theta - \theta^\prime)\,-\,1\,\right].
\label{covepsilon}
\eeq
One readily finds the value of the constant $D$ by equating the
mean square of $\tau_{m^\prime}-\tau_m$ obtained from (\ref{ddell})-(\ref{xdell})
to the mean square of $\tau(\phi^{\prime})-\tau(\phi)$ obtained from (\ref{rectautheta})-(\ref{covepsilon}) with $\ell = m^\prime-m$.
The result is that $D = 40\pi^2/27$.

The pair of equations (\ref{rectautheta})-(\ref{covepsilon}) is consistent with the scaling $\tau \sim n^{-3/2}$ found in equation (\ref{sctau}).
Of course we may scale $n$  out of (\ref{rectautheta})-(\ref{covepsilon}) by setting ${\tau}=n^{-3/2}\tilde{\tau} $ and ${\epsilon}=n^{-3/2}\tilde{\epsilon}$. 

The noise $\epsilon(\theta)$ would be white if it were not for the extra term $-1$ in the brackets in (\ref{covepsilon}), which adds a
coloring, even though an admittedly trivial one.
Equation (\ref{covepsilon}) is simpler than Langevin equations found in related work: The problem dealing with Sylvester's question \cite{Pfiefer89} and the problem of the $n$-sided Poisson-Voronoi cell without the Gabriel condition \cite{Hilhorst05b} are both associated with second order Langevin equations, the first one having trivial, and the second one nontrivial colored noise. 


\section{Conclusion}
\label{sec:conclusion}

We have derived an expression, valid in the limit of large $n$,
for the probability $\pnG$ that a planar Poisson-Voronoi cell be $n$-sided
while satisfying the additional condition of having only Gabriel neighbors.
We have contrasted $\pnG$ with the probability $p_n$, calculated in earlier work \cite{Hilhorst05b}, that a cell be $n$-sided but {\it without\,} this additional condition.
In both cases the initial expression has the form of a $2n$-dimensional phase space integral, similar to the partition function of $n$ interacting two-dimensional particles; and in both cases the key initial step is to transform to an integration over a set of angular variables.
The expressions for $p_n$ and $\pnG$ then differ by the limits of integration,
and from there on the calculation $\pnG$
proceeds in a different way.
Scaling arguments led us to neglect all terms of $\calO(1)$ in the asymptotic expansion of $\log\pnG$. The $2n$ angular integrations then decouple into $n$ pairs, subject only to two global constraints
represented by the two delta functions in (\ref{pnG2}).
One constraint is linear and the other one is nonlinear in the angles.
The corresponding constraints reappear in expression (\ref{pnG5}),
where it becomes clear that they
can be treated by integral representations with appropriate scalings with $n$. 

We have, subsequently, refined our analysis and explicitly calculated the joint probability distribution of a pair of angles $(\beta_m,\gamma_m)$, valid in the large $n$ limit. Such a pair is a local stochastic variable and in the limit of large $n$ the effect of the two constraints on its distribution vanishes.

In the problem without the Gabriel condition \cite{Hilhorst05b}
the deviations of the cell perimeter from circularity
contributed a constant $C$ to the prefactor of $p_n$.
Here, however, there is no such contribution to $\pnG$: the perimeter of an $n$-sided cell obeying the Gabriel condition is more rigidly circular than the perimeter of the unconstrained cell. This intuitively appealing effect has been quantified in the final section of this work.

We believe that this note constitutes an interesting addition to the general class of random geometry problems outlined in the introduction.


\end{document}